\documentclass[global]{svjour}
\usepackage{graphics}
\usepackage{color}
\journalname{Space Science Reviews}
\begin{document}
\title{ISO's Contribution to the Study of Clusters of Galaxies}
\author{ L. Metcalfe\inst{1} \and D. Fadda\inst{2} \and A. Biviano\inst{3}
}                     
\offprints{Leo.Metcalfe@sciops.esa.int}          
\institute{XMM-Newton Science Operations Centre, European Space Agency, 
Villafranca del Castillo, PO Box 50727, 28080 Madrid, Spain 
\and Spitzer Science Center, California Institute of Technology, 
Mail code 220-6, 1200 East California Boulevard, Pasadena, CA 91125
\and INAF - Osservatorio Astronomico di Trieste, via G.B. Tiepolo 11,
34131, Trieste, Italy}
\date{Received: date / Revised version: date}
%
\maketitle
\begin{abstract}
Starting with nearby galaxy clusters like Virgo and Coma, and
continuing  out to the furthest galaxy clusters for which ISO
results have yet been published ($z=0.56$), we discuss the
development of knowledge of the infrared and associated  physical
properties of galaxy clusters from early IRAS observations, through
the  ``ISO-era" to the present, in order to explore the status of
ISO's contribution to this field.  Relevant IRAS and ISO
programmes  are reviewed, addressing both the cluster galaxies and
the still-very-limited evidence for an infrared-emitting
intra-cluster  medium.

ISO made important advances in knowledge of both nearby and
distant  galaxy clusters, such as the discovery of a major cold
dust component in Virgo and Coma cluster galaxies, the elaboration
of the correlation between dust emission and Hubble-type, and the
detection  of numerous Luminous Infrared Galaxies (LIRGs) in
several distant clusters. These and consequent achievements are 
underlined and described.

We recall that, due to observing time constraints, ISO's coverage of
higher-redshift  galaxy clusters to the depths required to detect
and study statistically significant  samples of cluster galaxies
over a range of morphological types could not be comprehensive and
systematic, and such systematic coverage of distant clusters will
be an important achievement of the Spitzer Observatory.

\end{abstract}
\section{Introduction}
\label{sec:intro}
\subsection{ISO looks deep}
The most strongly star forming galaxies are heavily dust obscured,
and estimates of their star formation rates (SFR) made at visual 
wavelengths often fall one or two orders of magnitude below their
true values. Frequently, very actively star forming galaxies occur in 
associations or groups.  At the same time it has been believed that
in the dense environments of galaxy clusters the interactions of
galaxies with each other, with the cluster tidal field, and with the 
intra-cluster medium (via ram pressure) strip galaxies of their 
reserves of gas, and eventually suppress star formation.


Much has already been learned about the evolution of the cosmic
star formation rate in field galaxies from observations with
the European Space Agency's Infrared Space Observatory (ISO) satellite 
(Kessler et al.\,1996).
Thanks to deep surveys in the mid-infrared (MIR) and far-infrared (FIR) 
(e.g. Elbaz et al. 1999, 2002; Serjeant et al. 2000, 2004;
Gruppioni et al. 2002;  Lari et al. 2001; Metcalfe et al. 2003;
Sato et al. 2003; Rodighiero et al. 2003; Kawara et al. 2004; 
Rowan-Robinson et al. 2004; Fadda et al. 2004, and several others) conducted,
respectively, with ISOCAM\footnote{Throughout this paper the ISOCAM filters
having reference wavelengths 6.75 and 14.3\,$\mu$m will respectively
be referred to as the 7\,$\mu$m and 15\,$\mu$m filters. Both bandpasses
are referred to as mid-infrared\,(MIR).} (Cesarsky et al. 1996) 
and ISOPHOT (Lemke et al. 1996), we now know that the comoving
density of IR-bright galaxies has a very rapid evolution from $z
\sim 0$ to $z \sim 1$. This evolution has been interpreted as the
result of an increased rate of galaxy-galaxy interactions, coupled
with an increase in the gas content of the galaxies (Franceschini et al.
2001).
Dust obscuration plays an important role in concealing star formation
in field galaxies. Its role in relation to the star formation activity
of cluster galaxies is less known. Published ISO results to--date addressing the fields of 
galaxy clusters (Duc et al. 2002, 2004; Coia et al. 2004a \& b; Biviano et al. 2004; 
Metcalfe et al. 2003; Fadda et al.\,2000; Altieri et al. 1999; Soucail et al. 1999; Barvainis et al. 1999; 
Quillen et al. 1999; L\'emonon et al. 1998; Pierre et al. 1996.)
concern a dozen clusters at $z < 0.6$, and point to an important role for dust,
evolving with redshift, also in cluster galaxies.
The fraction of IR-bright, star-forming
cluster galaxies changes significantly from cluster to cluster,
without a straightforward correlation with either the redshift, or the
main cluster properties (such as the cluster mass and luminosity).

The presence of dust in cluster galaxies may have 
hampered our efforts to fully understand the evolution
of galaxies in clusters.  Optical-band observations have provided a
few, but very fundamental results, which cannot yet however be combined
in a unique, well constrained scenario of cluster galaxy evolution.

\subsection{Some properties of cluster galaxies}

Perhaps the most fundamental phenomenology that all scenarios of
cluster galaxy evolution (e.g. Dressler 2004) must explain is the
so called morphology density relation (hereafter MDR; Dressler
1980), whereby early-type galaxies, i.e. ellipticals and
S0s, dominate rich clusters, while late-type galaxies, i.e. spirals
and irregulars, are more common in the field. The fact that
early-type galaxies seem to reside in the cluster centres since
$z\,\geq\,1$--2, while the colour-magnitude relation (CMR,
Visvanathan \& Sandage 1977) remains very tight even at
$z\,\sim\,1$ suggests that the MDR is established at the formation
of galaxy clusters and that the early-type galaxies defining the CMR
are uniformly old and passively evolving since their formation
redshift,  $z_f>2$ (e.g. Ellis et al. 1997).
Similar conclusions are obtained by analysing the fundamental
plane (FP; Dressler et al. 1987), relating basic properties of
early-type galaxies (their effective radius, internal velocity
dispersion, and effective surface brightness). The FP, like the CMR,
still holds for early-type galaxies in $z\,\sim\,1$ clusters, and its
scatter is similar to that seen in nearby clusters (van Dokkum \&
Stanford 2003).

However, these conclusions could only be true {\em on average.} 
Independent analyses suggest that at least part of the cluster
galaxies have undergone significant evolution over the last 3--8 Gyr. First
and foremost is the observational evidence for an increasing fraction
of blue cluster galaxies with redshift, the so called `Butcher-Oemler'
(BO) effect (Butcher \& Oemler 1978, 1984;
Margoniner et al. 2001).
Approximately 80\% of galaxies in the cores of nearby clusters are
ellipticals or S0s, i.e. red galaxies (Dressler 1980), but the
fraction of blue galaxies increases with redshift. These blue
galaxies are typically disk systems with ongoing star formation 
(Lavery \& Henry 1988), with spectra characterized by strong
Balmer lines in absorption (typically, EW(H$\delta$) $> 3$ \AA) and no
emission lines, and have been named `E+A' (or also `k+a') galaxies 
(Dressler \& Gunn 1983).
Modelling of their spectra indicates that star formation stopped
typically between 0.05 and 1.5 Gyr before the epoch of observation,
in some cases after a starburst event (Poggianti et al. 1999, 2001).

Similarly, the fraction of spirals increases with redshift (Dressler
et al. 1997; Fasano et al. 2000; van Dokkum et al. 2001), at the
expense of the fraction of S0's.
Maybe, the colour and the morphological 
evolution of the cluster galaxy population are two aspects of the same
phenomenon. Field spirals are being accreted by clusters (Tully \& Shaya 1984;
Biviano \& Katgert 2004), and the accretion rate was higher in
the past (Ellingson et al 2001). It is therefore tempting to identify the
blue galaxies responsible for the BO-effect in distant clusters with
the recently accreted field spirals. 

Possibly, these evolutionary trends could be reconciled with the
passive evolution inferred from the CMR and FP studies by taking into
account the so-called `progenitor bias' (van Dokkum et al. 2000),
namely the fact that
only the most evolved among cluster early-type galaxies are indeed 
selected in studies of the colour-magnitude and fundamental plane 
relations.

\subsection{A hostile environment}

There is no shortage of plausible physical mechanisms that could drive
the evolution of a galaxy in a hostile cluster environment.  Among
these, the most popular today are ram pressure, collisions,
and starvation from tidal stripping (Dressler 2004), all in principle
capable of depleting
a spiral of its gas reservoir, thereby making it redder and more
similar to a local S0.

Ram pressure from the dense intra-cluster medium can sweep cold gas
out of the galaxy stellar disk (Gunn \& Gott 1972) and induce star 
formation via compression of the gas that remains bound to the
galaxy. Collisions or close encounters between galaxies generate
tidal forces that tend to funnel gas towards the galaxy centre (Barnes
\& Hernquist 1991) eventually fueling a starburst that ejects
gas from the galaxy. The cumulative effect of many 
minor collisions (named `harassment', Moore et al. 1996), can lead 
to the total disruption of low surface brightness galaxies (Martin 1999). 
The collision
of a group with a cluster can also trigger starbursts in cluster
galaxies, as a consequence of the rapidly varying tidal
field (Bekki 2001). Finally, the so called `starvation' mechanism
(Larson et al. 1980) affects the properties of a galaxy by simply
cutting off its gaseous halo reservoir. This can occur because of
tidal stripping, a mechanism effective in galaxy-galaxy encounters,
but also when galaxies pass through the deep gravitational potential
well of their cluster. 

The common outcome of all these processes is galaxy gas depletion,
ultimately leading to a decrease of the star formation activity for
lack of fuel, and, hence, to a reddening of the galaxy stellar
population. However, some of these processes induce a starburst phase
before the gas depletion, and some do not.

To date, it remains unclear which physical process dominates in the
cluster environment. Useful constraints can be obtained by finding
{\em where} the properties of cluster galaxies change with respect
to the field, since the different processes become effective at
different galaxy or gas densities. Recently it has been
found (Kodama et al. 2001; G\'omez et al. 2003; Lewis et al. 2002)
that a major change in the star-formation properties of cluster
galaxies occurs in the outskirts of clusters (at $\sim\,1.5$ cluster
virial radii).  These results would seem to exclude ram-pressure
stripping as a major factor in cluster galaxy evolution, since the
density of the intra-cluster medium is too low in the cluster
outskirts.

Recent observations of high-$z$ clusters seem to have complicated,
rather than simplified, the issue of cluster galaxy evolution.  A
surprisingly high fraction of {\em red} merger systems has been found
in these distant clusters (van Dokkum et al. 2001).
The red colours of these merging galaxies and the lack of emission
lines in their spectra suggest that their stellar populations were
formed well before the merger events, but the occurrence of relatively
recent starburst events in these galaxies is instead suggested by
detailed analyses of their spectra (Rosati 2004).

\subsection{An uncluttered view}

A better understanding of the evolutionary processes affecting cluster
galaxies can come from observations in the infrared. Dust, if present, is
capable of obscuring most of a galaxy's stellar radiation, making the
observed galaxy red and dim at optical wavelengths, and affecting
optical estimates of the galaxy star formation activity.  The effects of
dust can be particularly severe if the galaxy is undergoing a
starburst (Silva et al.  1998), so that we might be missing a
substantial part of the evolutionary
history of cluster galaxies by observing them at optical wavelengths.
Since the dust-reprocessed stellar radiation is re-emitted at IR
wavelengths, the IR luminosity is a much more reliable indicator of a
galaxy's star formation activity (Elbaz et al. 2002).

The plan of this review is as follows: in Section 2 the development of 
knowledge of the infrared properties of galaxy clusters from early IRAS
observations, through the ``ISO-era" to the present is described. Section 2.1 
considers the accumulation of data on the Virgo cluster, while 
Section 2.2 addresses other nearby clusters, e.g. the Fornax, Hydra, Coma and 
Hercules clusters. Section 2.3
discusses the significant progress that has been possible with ISO
in the study of cluster galaxy properties out to moderate redshifts
($<$\,0.6) and attempts to draw some comparison among the still rather 
heterogeneous sample of cluster observations. Section 2.4 reviews
the status of attempts to directly observe diffuse intra-cluster dust in the
infrared. Finally, Section 3 summarises the current status of the field
and remarks on the important opportunity represented by the Spitzer
Observatory to decisively extend the field.

\section{Cluster galaxies in the infrared}
\label{sec:Clusters}
\subsection{The Virgo galaxy cluster}

Being the most nearby relatively massive galaxy cluster, the Virgo cluster
has been studied extensively at all wavelengths. Already in 1983, 
Scoville et al. obtained 10\,$\mu$m data with IRTF
for 53 Virgo spiral galaxies and concluded that star formation is
occurring in the nuclei of virtually all spiral galaxies independent
of spectral type, with an average star formation rate (SFR) of 
0.1\,M$_{\odot}$/yr.  No correlation was found between 10\,$\mu$m emission
and gross properties (barred or normal, early or late spiral
morphology, total optical luminosity), nor with location in the
cluster. Using far- to mid-infrared flux correlations, they inferred a 
far-infrared\,(FIR)
luminosity $\sim 2\,\times\,10^{10}\,L_{\odot}$ for the brightest
10\,$\mu$m source in the Virgo cluster, NGC4388.

Leggett et al. (1987) made optical identifications of 145 IRAS
galaxies in the 113 square degree field centred on the Virgo
cluster and concluded that they were mostly seeing the spirals, and
that the infrared properties of the Virgo cluster galaxies
are indistinguishable from those of field galaxies at similar
redshift. Virgo spirals were indistinguishable from
field disc galaxies with normal SFRs. The typical infrared 
luminosity for the sample was $L_{IR}\,\sim\, 10^9L_{\odot}$, and 
the most luminous confirmed cluster sources were found to have 
$L_{IR}\,=$ a few $\times\,10^{10}L_{\odot}$.  They concluded that 
the cluster environment has no effect on galaxy IR properties, even 
for very HI
deficient galaxies. Their conclusion was however criticized by Doyon
\& Joseph (1989) who, using a sample of 102 Virgo spirals detected at
60 and 100\,$\mu$m, were able to show that HI-deficient galaxies have
lower IR fluxes, lower star formation activity, and cooler IR colour
temperatures than those with normal HI content.

If cluster environment affects the interstellar medium (ISM) of
cluster galaxies, one might expect the FIR, radio and FIR-radio
correlation to be affected.  Niklas et al. (1995) looked at the
FIR-radio correlation in Virgo galaxies and suggested that, in this
respect, most of the Virgo galaxies behave like normal field
galaxies. Only a few early type spirals with strong central sources as
well as very disturbed galaxies show a high radio excess. These are in
the inner part of the cluster, where galaxies have disturbed
HI-distributions, and truncated HI-disks, as shown by 21\,cm
observations (Cayatte et al. 1990). Measurements of molecular gas
emission in radio continuum (Kenney \& Young 1986, 1989) show a much
less pronounced stripping effect.

In the ISO era, Tuffs et al. (2002) and Popescu et al. (2002a, b)
used ISOPHOT at 60, 100 and 170\,$\mu$m, to study a luminosity- and
volume-limited sample of 63 S0a or later-type galaxies in the core and
periphery of the Virgo cluster. They reached sensitivities 10 times
better than IRAS in the two shorter-wavelength bandpasses, and the
confusion limit at 170 $\mu$m. These programmes sought to extend
knowledge of FIR SEDs to lower limits covering a complete sample of
normal\footnote{A `Normal' galaxy is understood to mean a galaxy not
dominated by an active nucleus, with SFR sustainable for a substantial
fraction of a Hubble time.} late-type galaxies over a range of
morphological type and star formation activity.

A significant cold dust component (with a temperature of around 18~K)
was found in all morphological classes of late-type galaxies, from
early giant spirals to irregular galaxies and Blue Compact Dwarfs
(BCDs), and which could not have been recognized by IRAS. 
These results required a revision of the masses and
temperatures of dust in galaxies. On average, dust masses are raised
by factors of 6 to 13 with respect to IRAS results. 
The FIR/radio correlation is confirmed for the warm
FIR dust, and is found for the cold dust. The predominance of the very
cold dust component (down to less than 10\,K) in BCDs was remarkable,
with a few 10s of percent of the UV/Optical component appearing in the
cold dust emission. The cold emission might be due to collisional dust
heating in dust swept up in the intergalactic medium (proto-galactic
cloud) by a galactic wind from the BCD or, alternatively, to photon
heating of the dust particles in an optically thick disk, indicative
of a massive gas/dust accretion phase that makes BCDs sporadically
bright optical/UV sources when viewed out of the disk equatorial
plane.

On average, 30\% of the stellar light of spirals in Virgo is
re-radiated by dust, with a strong dependence on morphological type,
ranging from 15\% for early spirals to 50\% for some late spirals, and
even more for some BCDs (Popescu \& Tuffs 2002). Fig.\,1, taken from 
Popescu \& Tuffs (2002), shows the dependency of the dust contribution 
on the Hubble type in the Virgo cluster, and it illustrates a sequence 
of increasing FIR-to-total bolometric output running from normal
to gas-rich-dwarf galaxies.

\begin{figure}
\resizebox{0.5\hsize}{!}{\includegraphics{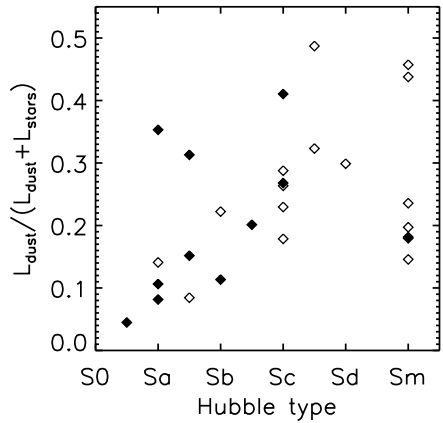}}
\caption{From Popescu \& Tuffs (2002): The ratio of dust luminosity
to total bolometric luminosity as a function of Hubble Type, for Virgo cluster
galaxies. Filled symbols are cluster core galaxies and open symbols refer to the cluster periphery. 
A correlation is established between Hubble type and the FIR contribution to bolometric luminosity.}
\end{figure}

Leech et al. (1999) studied a sample of 19 Virgo cluster spiral
galaxies with ISO's Long Wavelength Spectrometer (LWS) (Clegg et al. 1996) obtaining
spectra around the [CII] 157.7\,$\mu$m fine-structure line for 14 of
them. [CII] line radiation provides the most important gas cooling
mechanism in normal, i.e. non-starburst, late-type galaxies, balancing
photoelectric heating from grains.  In field galaxies [CII] is typically
10$^{-3}$ to 10$^{-2}$ of total galaxy FIR. The sample, drawn from
both the cluster core and the cluster periphery (and being a
sub-sample of the Tuffs et al. 2002 sample), spanned the S0a--Sc
morphological range to probe any difference in the [CII] emission
between different types or between core and periphery galaxies.

A good correlation was found between the strength of the [CII] line
and the FIR flux, as measured by IRAS. Moreover, the
[CII]-to-$K'$-band flux ratio shows a two order of magnitude
difference between early-type galaxies and late spiral types. Galaxies with large
[CII]/FIR ratios tend to have later Hubble types. No apparent
relationship was found between [CII] strength and galaxy position in
the cluster, nor between [CII] and HI-mass surface density. Any
influence of the Virgo cluster environment on the [CII] emission was
found to be small compared with the strong dependence of the line
emission on basic measurables such as morphology or bulk mass of the
stellar component, as measured by the near-IR ($K'$-band) luminosity.

In a series of papers, Boselli et al. (1997a,b, 1998, 2003a,b, 2004)
explored the properties of a large sample of ISO-detected spiral and
irregular galaxies in the Virgo cluster, in measurements made with
ISOCAM at 7 and 15\,$\mu$m.  71 objects were in the cluster periphery
($\geq$\,4 degrees from M87) and 28 in the core ($\leq$\,2 degrees
from M87) in order to allow study of the effects of the environment on
dust emission and evolution. A further 24 Virgo cluster galaxies were
serendipitously observed, bringing the full Virgo sample up to 123
galaxies. S0 and elliptical galaxies observed by chance in the ISO
fields were used to establish the stellar contribution to the IR
emission. Objects had luminosities in the range
$10^{7.4}\,\leq\,L_{FIR}\leq\,10^{10.1}\,L_{\odot}$.

Thirty four of the Virgo objects were fully resolved by ISOCAM, and
MIR images could be presented for these, along with radial light and
colour profiles and other morphological and structural information.

Boselli et al. showed that the MIR emission of optically-selected,
normal early-type galaxies is dominated by the Rayleigh-Jeans tail of
the cold stellar component, while that of late-type galaxies is
dominated by the thermal emission from dust, but partly contaminated
by stellar emission, especially at 7\,$\mu$m in early-type spirals
(Sa). While the MIR emission (per unit mass) of the spirals and later
systems are comparable, the average 7\,$\mu$m and 15\,$\mu$m to
$K'$ flux ratio of E, S0 and S0/a galaxies is significantly lower than
that of spirals. At 15\,$\mu$m, where the difference is clearer,
spirals have on average a MIR emission per unit mass higher by more
than one order of magnitude than E-S0/a. BCDs have, on average, a MIR
emission per unit mass comparable to that of spirals.

The IR emission carriers and their behaviour are consistent with
quantitative expectations based on MIR studies of the ISM in our
Galaxy.  In spiral and irregular galaxies the 7\,$\mu$m emission is
almost entirely due to the UIB\footnote{The Unidentified Infrared Bands (UIB)
dominate the 5 to 12 micron MIR spectrum of a wide range of celestial
sources, and are usually assumed to arise from polycyclic aromatic
hydrocarbon molecules (PAHs).} carriers.  The MIR fluxes are
proportional to the SFR when it is not too large, but fall off in
the presence of a high SFR suggesting that the UIBs are destroyed
by the UV field. However, in galaxies with a high SFR, there is an
additional diffuse contribution (i.e. not well correlated with HII regions or H$_{\alpha}$
sources) to the 15\,$\mu$m flux from very small,
three-dimensional, grains.  As a consequence, MIR dust emission is not an optimal
tracer of star formation in normal, late-type galaxies, and MIR
luminosities are better correlated with FIR luminosities than with
more direct tracers of the young stellar population such as the
H$\alpha$ and the UV luminosity. This conclusion, valid for nearby
normal, late-type galaxies, may not apply to luminous starburst galaxies
(Luminous Infrared Galaxies, or LIRGs), such as those detected in the 
ISO deep surveys (F\"orster-Schreiber et al. 2004). 
The MIR emission traces well the FIR and bolometric emission
(Boselli et al. 1998; Elbaz et al. 2002).

\subsection{The Coma, Fornax, Hydra, Hercules \& other nearby ($z\,<\,0.1$) clusters}
\label{sec:Coma}
Located at much larger distance than Virgo, but with a much larger mass,
Coma has often been a favourite observational target, also in the IR.

Wang et al. (1991) made optical identifications of a total of 231
IRAS point sources in the regions of the Fornax, Hydra and Coma
clusters, and identified respectively 13, 29 and 26 cluster galaxies.
They concluded that the cluster environment has no detectable
influence on galaxy infrared properties. Bicay \& Giovanelli (1987) came
to the same conclusion, based on a sample of 200 FIR-emitting galaxies
in seven nearby clusters (including Coma). However, they remarked that
LIRGs were much less common in the clusters than in the field (see
also Section~\ref{sec:general}).

The most distant cluster studied with IRAS was A2151 (aka Hercules, at
$z=0.036$). Out of 41 sources detected at 60$\mu$m in a 1.6$\times$2.5
sq.deg. field, Young et al. (1984) correlated 24 with late-type spiral
galaxies of the cluster remarking the total absence of IR emission
from E and S0 galaxies. Odenwald (1986) found CO emission in 3 of the 
9 most optically luminous galaxies in the spiral-rich Ursa
Major I(S) galaxy group, seeking evidence for galaxy interactions in a
cluster lacking an intra-cluster medium, but found little evidence that
processes unrelated to the galaxies themselves had influenced their
histories. Studying a sample of 200 galaxies in seven
nearby clusters\footnote{A262, Cancer, A1367, A1656
(Coma), A2147, A2151 (Hercules), and Pegasus}, Bicay \& Giovanelli (1987) 
pointed out the absence of
luminous IR galaxies (LIRGs, i.e. galaxies with $L_{FIR}\,>\,10^{11}\,
L_{\odot}$).  The sample consisted almost entirely of IR normal
galaxies ($L_{FIR}\,<\,10^{10}\,L_{\odot}$) in contrast to the rather high
percentage of LIRGs (20\%) detected by IRAS in the field.  Moreover,
the lack of a strong correlation between galaxy HI content and IR
emission led them to conclude that SFR is not enhanced by interaction with 
the ICM, and might even be quenched by it.  On the contrary, the suppressed
FIR-to-radio ratio of spiral galaxies found in rich clusters with
respect to poor clusters (Andersen \& Owen, 1995) seemed to suggest
that ram pressure enhances the radio emission in rich clusters while
galaxy-galaxy interactions play  a more important role in poor clusters
where velocity dispersion, and so encounter velocities, are smaller.

Quillen et
al. (1999) observed 7 E+A galaxies plus one emission-line galaxy at
12\,$\mu$m with ISOCAM. They found that E+A galaxies have mid- to
near-IR flux ratios typical of early-type quiescent galaxies, while
the emission-line galaxy had enhanced 12\,$\mu$m emission relative to
the near-IR. Galaxies with ongoing star formation have a different velocity
distribution in the cluster from galaxies with stopped SF, suggesting
that the ongoing infall of field spirals into the cluster potential
may first trigger and then quench star formation.

Further observations of Coma cluster galaxies in the MIR came from
Boselli et al. (1998) and Contursi et al. (2001). They also observed
the cluster A1367, located in the Coma supercluster, detecting, in total, 
18 spiral/irregular galaxies in the MIR and FIR with ISO.

Confirming results found in Virgo galaxies, these authors concluded
that most IR-detected Coma galaxies display diffuse MIR emission
unrelated to their H$\alpha$ emission. The aromatic carriers are not
only excited by UV photons, but also by visible photons from the
general ISM. When the UV radiation field is too intense, it can even
destroy the aromatic carriers, and overall the MIR emission is
dominated by photo-dissociation regions rather than HII-like regions.
A cold dust component was detected in all galaxies, at temperature of
$\sim 22$~K, more extended than the warm dust.
Only a very weak trend was found between the total dust mass and the gas
content of the galaxies, even if some galaxies are very HI-deficient,
and there was no detection of any relation between the MIR/FIR
properties and the environment.

All these results seemed to suggest very little (if any) dependence of
the IR properties of galaxies on the environment. If anything, IR
emission was thought to be quenched in the cluster
environment. However, only nearby clusters had been studied, in which
most of the galaxies are early type with little gas (and
dust).  With the launch of ISO it became possible for the first time
to study galaxy clusters out to redshifts where a significant change
in the composition of the cluster galaxy population had already been
(Butcher \& Oemler 1984) or was soon to be (Dressler et al. 1999)
observed in the optical.

\subsection{Galaxy clusters at intermediate redshift}
\label{sec:general}

ISO's mid-infrared camera, ISOCAM, with its vastly
improved sensitivity and spatial resolution with respect to IRAS,
has successfully observed several galaxy clusters out to redshifts
at which significant evolution might be expected to occur. 
At the same time, while 
studies of galaxies in nearby clusters are frequently done by targeting galaxies 
selected at other wavelengths, distant clusters can be completely 
surveyed due to their smaller angular size, producing an unbiased 
sample of infrared-emitting galaxies.
Published ISO observations to date for clusters at redshifts above
0.1 (Coia et al.\,2004a, b; Biviano et al.\,2004; Metcalfe et
al.\,2003; Duc et al.\,2002, 2004; Fadda et al.\,2000; Barvainis et
al.\,1999;  Altieri et al.\,1999; L\'emonon et al.\,1998 and Pierre
et al.\,1996) address seven 
clusters (see Table.\,1)
spanning the redshift range $0.17\,<\,z\,<\,0.6$ and yield
MIR data for around 110 cluster galaxies, slightly over 40 of these
seen at 15\,$\mu$m, and the rest only in the 7\,$\mu$m bandpass.

\begin{table*}[ht2\columnwidth]     
\caption[]{Clusters in the redshift range $0.17\,<\,z\,<\,0.6$ studied
with ISOCAM. The content of the columns in the table is as
follows: name and redshift of the cluster; number, respectively, of 15\,$\mu$m-only,
7\,$\mu$m-only, and 7-and-15\,$\mu$m confirmed cluster sources detected in each case, and
total number of MIR sources detected.
Results are gathered from Coia et al.\,(2004a, b), Biviano et al. (2004), Metcalfe et al.\,(2003),
Duc et al.\,(2002, 2004) and Fadda et al.\,(2000).
\label{clusters-list}}
\begin{center}
\leavevmode \small
      \begin{tabular}{cccccc}
\hline
\hline
\noalign{\smallskip}
Cluster       &   z   &  No.             &  No.            &  No.              &  No. all \\
              &       &  15\,$\mu$m-0nly &  7\,$\mu$m-0nly &  7-and-15\,$\mu$m &  sources \\
              &       &  sources         &  sources        &  sources          &          \\
\hline \noalign{\smallskip}
A2218         & 0.175 &  5               &  18             &  4                &   27     \\
\noalign{\smallskip}
A1689         & 0.181 &  3               &  20             &  9                &   32     \\
\noalign{\smallskip}
A1732         & 0.193 &  -               &   4             &  0                &    4     \\
\noalign{\smallskip}
A2390         & 0.23  &  1               &  11             &  3                &   15     \\
\noalign{\smallskip}
A2219         & 0.228 &  3               &   -             &  -                &    3     \\
\noalign{\smallskip}
A370          & 0.37  &  1               &   5             &  0                &    6     \\
\noalign{\smallskip}
CL0024+1654   & 0.39  & 13               &   -             &  -                &   13     \\
\noalign{\smallskip} 
J1888.16CL    & 0.56  &  6               &   -             &  -                &    6     \\
      \hline
       \end{tabular}
\end{center}
\end{table*}

\begin{figure}
\resizebox{1.0\textwidth}{!}{\includegraphics{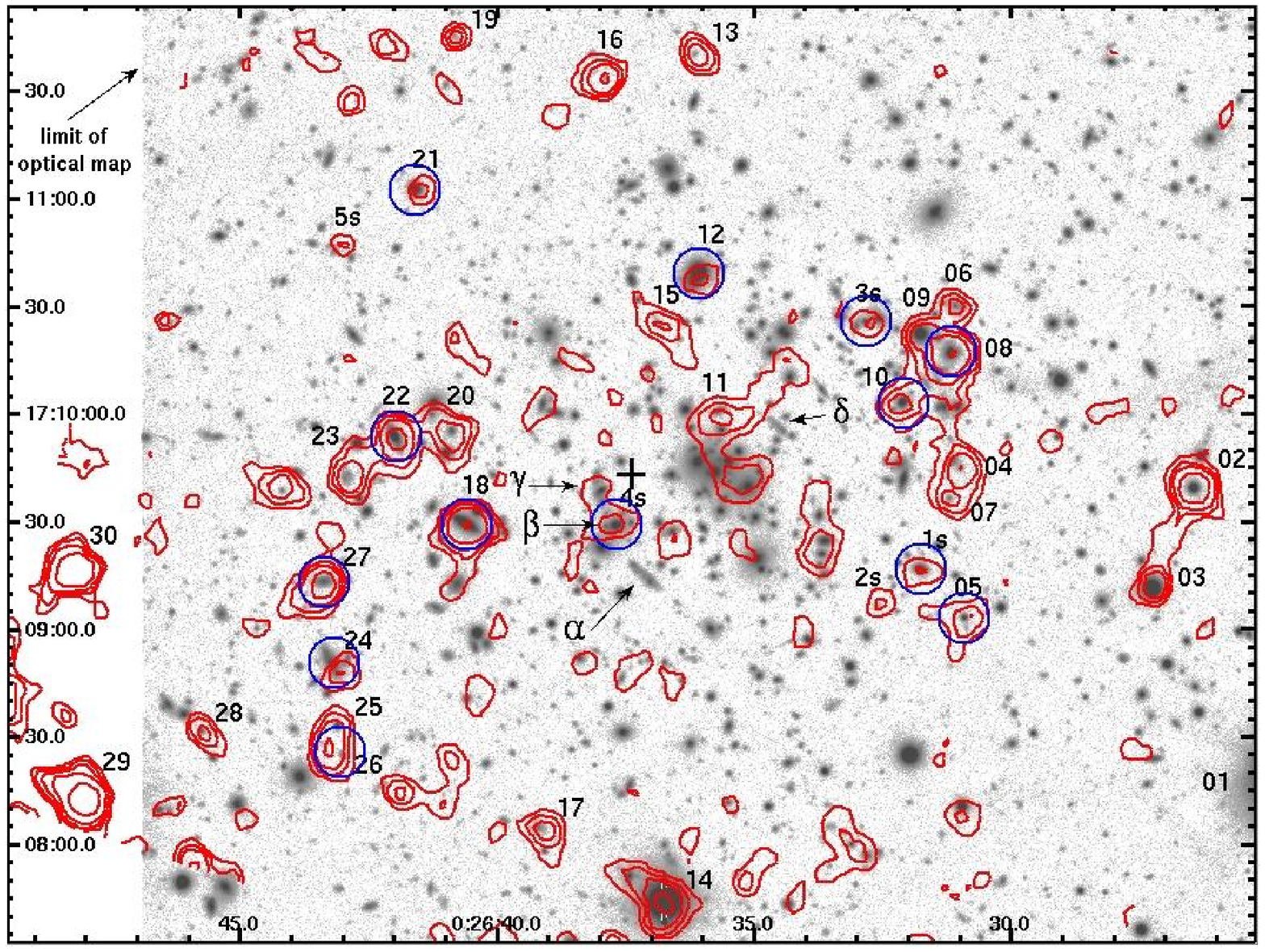}}
\vskip 0.5cm
\caption{From Coia et al. (2004b): Contours of a 15\,$\mu$m map of 
the (gravitationally lensing) galaxy cluster CL0024+1654 overlaid on a V-band FORS2 VLT image.  15\,$\mu$m
sources are numbered in order of increasing R.A. Dark circles identify
spectroscopically confirmed cluster galaxies. Greek letters denote four
prominent gravitational lensing arcs. North is up and East to the left,
with the centre of the map falling at R.A. 00 26 37.5 and DEC. 17 09 43.4 (J2000).}
\end{figure}

Almost half of the cluster galaxies detected at 15\,$\mu$m prove
to be LIRGs. (At the higher redshifts of the above sample only
LIRGs fall above the sensitivity limit of the observations.) About
60\% of these cluster galaxies were detected in observations 
originally intended to study distant field galaxies via the
gravitational lensing amplification of the foreground clusters
(Metcalfe et al.\,2003; Barvainis et al.\,1999; Altieri et al.\,1999),
and which, being generally very deep spatially-oversampled measurements, 
were able to provide insights
about the lensing cluster galaxy populations (Biviano et al. 2004;
Coia et al. 2004a and b).  Fig.\,2, taken from Coia et al. (2004b) 
is a V-band image of the $z\,=0.39\,$ galaxy cluster CL0024+1564 overlaid with 
contours of an ISO 15\,$\mu$m map.  The capacity of ISOCAM to detect 
and assign MIR counterparts unambiguously to numerous galaxies in 
the field is evident.

The first published ISOCAM observations of a distant cluster were those of the
$z=0.193$ cluster A1732 by Pierre et al.\,(1996), which they observed at 
7 and 15\,$\mu$m over an $8\,\times\,8$\,arcmin$^2$ field.  
They found some evidence for a deficiency of spirals and star forming
galaxies in the cluster, identifying only four cluster sources (at
7\,$\mu$m, no cluster sources were detected at 15\,$\mu$m) and 10 
MIR galaxies in total in the field, most of them judged to be foreground.  
Nevertheless, these were the faintest MIR extragalactic sources reported up to
that point and underlined the need for ultra-deep observations to detect
cluster members at 15\,$\mu$m. 

L\'emonon et al.\,(1998) reported evidence for an active
star-forming region in a cooling flow (later `cool-core') from 7
and 15\,$\mu$m observations of the inner square arcminute of the
well known lensing cluster A2390 ($z =0.23$), with an attendant
SFR of as much as 80\,$M_{\odot}\,yr^{-1}$ in the central cD. But 
this was later found to be compatible with non-thermal emission from a
jet associated with the cD (Edge et al. 1999). The estimated
cluster mass deposition rates in cooling flows have since been
lowered by one or two orders of magnitude (B\"ohringer et al. 2002).

A1689 ($z=0.181$) was the first distant cluster for which detailed
ISO observations were reported. Fadda et al.\,(2000) detected
numerous cluster members (30 at 7\,$\mu$m and 16 at 15\,$\mu$m) within
0.5~Mpc of the cluster centre, and they found a
correlation between the B-15\,$\mu$m colour and cluster-centric distance
of the galaxies. The 15\,$\mu$m galaxies are blue outliers with respect to 
the colour/magnitude relation for the cluster and become brighter
going from the center to the outer parts of the cluster. Coupled with the 
systematic excess of the
distribution of the B-15\,$\mu$m colours with respect to nearby
clusters (Virgo and Coma), this suggested the existence of an IR
analogue of the Butcher-Oemler effect in A1689 (see Fig.\,3).
A follow-up optical study of these
infrared galaxies (Duc et al. 2002) showed that the morphology of the
15\,$\mu$m sources in A1689 is generally spiral-like, with
disturbances reminiscent of tidal interactions. 
No LIRGs were found in A1689. The highest total IR luminosity found
for a cluster galaxy was $6.2\times 10^{10}L_{\odot}$, corresponding
to a SFR of $\approx 11\,M_{\odot}$ per year. The median SFR for
A1689, derived from 15\,$\mu$m measurements, was $2\,M_{\odot}$ per
year, while the median found from [OII] (optical) measurements was only
$0.2\,M_{\odot}$ per year. 

\begin{figure}
\resizebox{0.5\hsize}{!}{\includegraphics{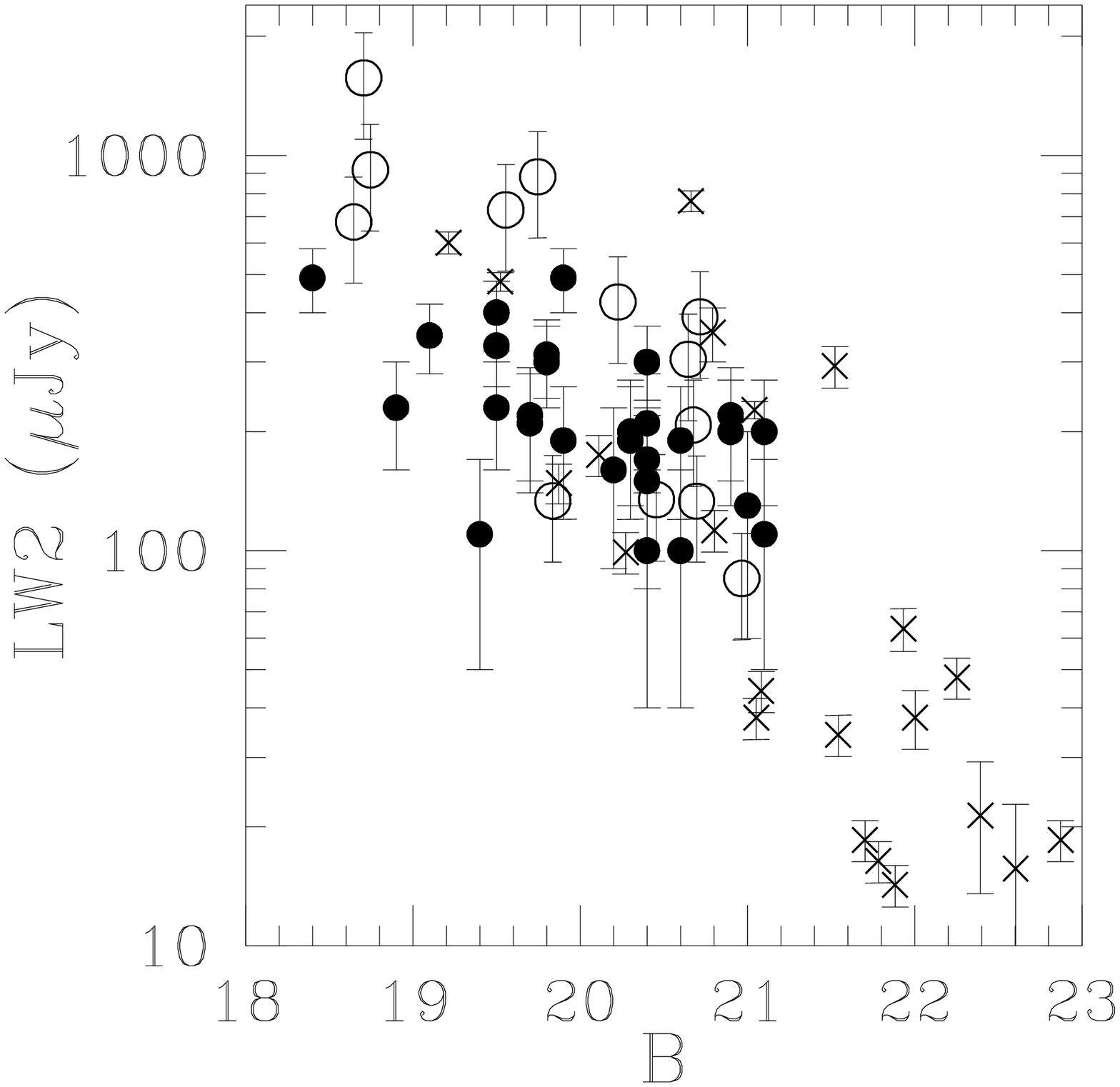}}
\resizebox{0.5\hsize}{!}{\includegraphics{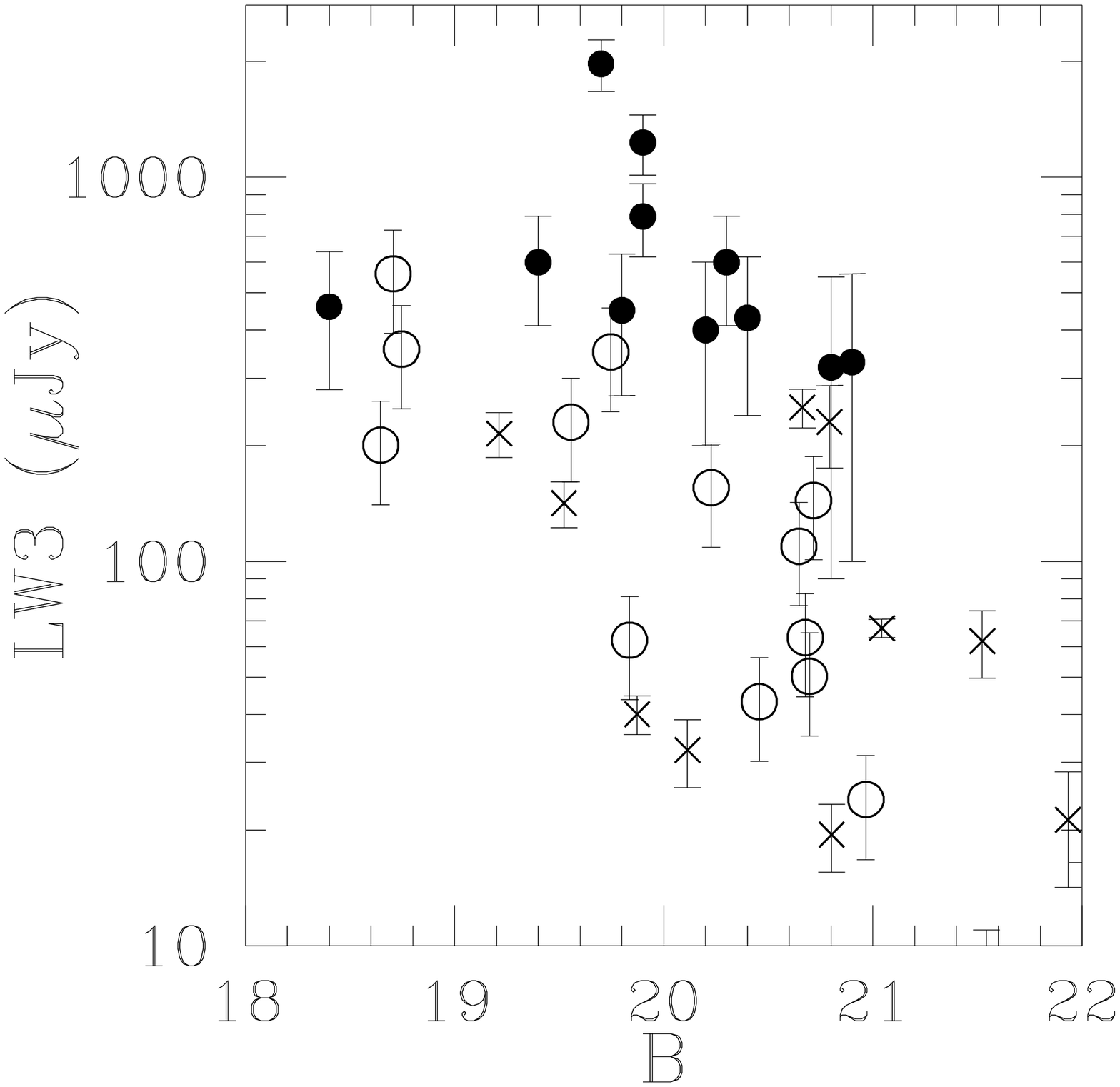}}
\caption{Fadda et al.\,(2000) found the [B$_T$\,-\,7\,$\mu$m] colour distribution of A1689 cluster galaxies
(filled circles) to
be compatible with that of the nearby Virgo (crosses) and Coma (empty circles) clusters, but the
[B$_T$\,-\,15\,$\mu$m] colour distribution showed a systematic excess
with respect to those nearer clusters.  The cluster
contains an excess of 15\,$\mu$m sources relative to the field,
suggesting that the environment of A1689 triggers starburst episodes
in galaxies in the cluster outskirts that have similar IR
luminosities and FIR/optical colours to those of field starburst
galaxies.}
\end{figure}

This paper revealed
the importance of IR observations in the study of star-formation in
clusters. About one third of the 15\,$\mu$m sources show no sign of
star formation in their optical spectra. Moreover, comparing the
star-formation estimates from IR and [OII] (see Fig.\,4), 
Duc et al.\,(2002) deduced that at least 90\% of the star formation 
activity taking place in A1689 is obscured by dust.

\begin{figure}
\resizebox{1.0\textwidth}{!}{\includegraphics{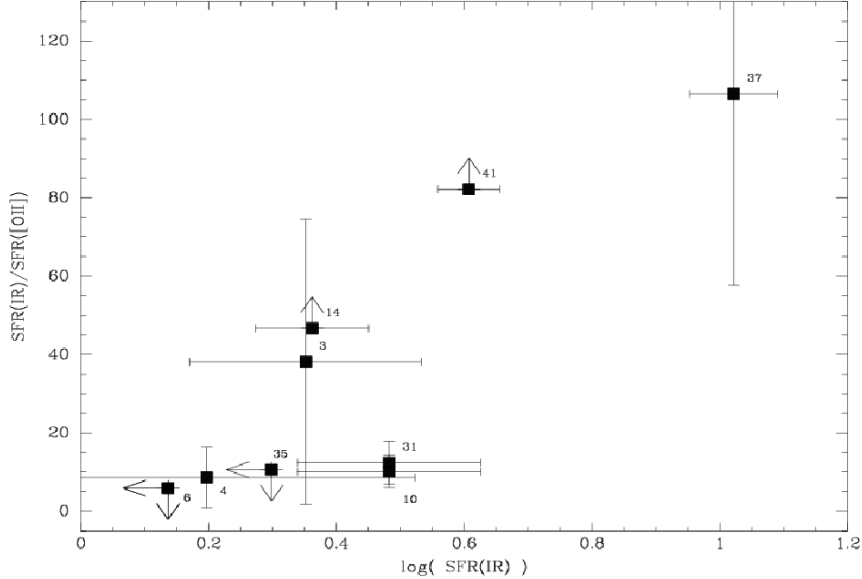}}
\vskip 0.5cm
\caption{This plot, taken from Duc et al.\,(2002), compares the 
A1689 SFR derived from optical [OII] measurements, and from ISO
MIR measurements. It illustrates the 
limitations of purely optical indicators of star formation rates. 
A major part, at least 90\%, of the star formation activity 
taking place in A1689 is hidden by dust at optical wavelengths.
}
\end{figure}

The ISO gravitational lensing survey programme (Metcalfe et al. 2003),
and related work, led to deep observations of the core of several 
distant clusters.  Large numbers of cluster galaxies were detected
in the fields of A2218, A2390 and CL0024+1654. So far, the cases of 
A2218 and CL0024+1654 have been treated in 
dedicated papers. 

In the analysis of the galaxies in the field of 
A2218, a rich cluster at $z=0.175$, Biviano et al.\,(2004) found 9 
cluster members at 15$\mu$m inside a radius of 0.4~Mpc. In contrast 
to the case of A1689, which is at almost identical redshift ($z=0.181$) 
and for which the
median SFR is about 2$M_{\odot}\,yr^{-1}$ and median $L_{IR}$ is 
around $10^{10}\,L_{\odot}$ , only one
of the A2218 MIR galaxies is a blue Butcher-Oemler galaxy. The
MIR luminosity of A2218 galaxies is moderate and the inferred star
formation rate is typically less than 1$M_{\odot}\,yr^{-1}$ with a median
$L_{IR}$ of only $6\,\times\,10^{8}\,L_{\odot}$.
The absence
of a MIR BO effect in A2218 might be a consequence of the small area 
observed, about 20.5 square arcminutes ($r < 0.4Mpc$), and yet the area 
studied for A1689 was not much larger, at about 36 square arcminutes. 
Coia et al. (2004b) suggest that the difference between these two clusters
may be traced to their having different dynamical status. 

The same sort of comparison can be drawn between the clusters 
CL0024+1654 ($z=0.39$) and A370 ($z=0.37$).  These two clusters at
similar redshift and mapped in almost identical ways, exhibit very
different numbers of LIRGs (Table 2),
probably, according to Coia et al., because an ongoing cluster merger 
gives rise to enhanced star-forming activity in CL0024+1654.

As can be appreciated from Fig.\,5 taken from Coia et al.\,(2004b), the median infrared luminosity of 
ISO-detected cluster galaxies in CL0024+1654 is around $1\,\times\,10^{11}\,L_\odot$.
Star formation rates derived from the 15\,$\mu$m data range from 8 to 77\,$M_{\odot}\,yr^{-1}$,
with median(mean) value of 18(30)\,$M_{\odot}\,yr^{-1}$.

\begin{figure}
\resizebox{0.5\hsize}{!}{\includegraphics{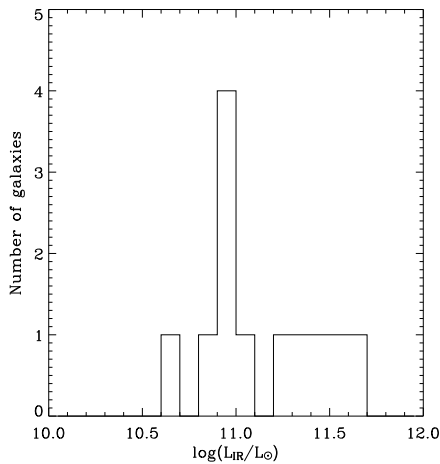}}
\caption{The infrared luminosity distribution for MIR galaxies in CL0024+1654. The median value
for L$_{IR}$ is $\sim\,1.0\,\times\,10^{11}\,L_{\odot}$. (From Coia et al. 2004b.)}
\end{figure}

Because of the different sky areas mapped with ISO for several of the clusters 
discussed here, and the different sensitivities achieved, it is not straightforward
to compare the results for different clusters.  A useful approach is to compare
the number of LIRGs detected in each cluster, since these IR-bright galaxies
would be expected to be seen for any of the observations considered.  Such 
a comparison, taken from Coia et al.\,2004b, is presented in Table 2.  For each cluster 
an ``expected" number of LIRGs is derived to test the hypothesis that the
cluster is similar to the LIRG-rich cluster CL0024+1654\footnote{In fact, to avoid throwing away 
several sources close to the LIRG flux threshold of $1\,\times\,10^{11}\,L_{\odot}$ and thereby 
degrading the statistics of the comparison, the flux threshold for the comparison
of detected luminous sources was set to $9\,\times\,10^{10}\,L_{\odot}$}. The LIRG count in 
CL0024+1654 is multiplied by the ratios of (a) virial mass per unit area of the cluster to 
that of CL0024+1654, the square of the respective distances to the cluster and to
CL0024+1654, and the observed solid-angle for the cluster to that of CL0024+1654.
The resulting column of the table can then be compared with the column listing the
actual observed number of LIRGs for each cluster. 

\begin{table*}[ht2\columnwidth]     
\caption[]{Summary of ISOCAM observations and results at 15 $\mu$m
for five clusters of galaxies.  The table is adapted from Coia et al.\,2004b, and
the data originates from Metcalfe
et al.\,(2003) for Abell 370, Abell 2218 and
Abell 2390, Fadda et al.\,(2000) and Duc et
al.\,(2002) for Abell 1689, and Coia et al.\,(2004) for
CL0024+1654.  The content of the columns in the table are as
follows: name and redshift of the cluster, total area scanned,
sensitivity reported at the $5\sigma$ level, flux of the weakest
reported source in $\mu$Jy. Then number of cluster galaxies, total number of
sources detected including sources without redshift and stars.
virial radius of the cluster, virial mass, number
of sources with $\mathrm{L_{IR}>9\,\times\,10^10\,L_{\odot}}$ detected and
expected. The expected number of sources was obtained by comparison
with CL0024+1654 as described in the text. Virial radii and masses
are from Girardi \& Mezzetti\,(2001) and King et al.\,(2002). 
\label{comp}}
\begin{center}
\leavevmode \small
      \begin{tabular}{ccclrrrccccc}
\hline
         \hline
\noalign{\smallskip}
Cluster& z    & Area        &   \multicolumn{2}{c}{Sensitivity}   & \multicolumn{2}{c}{{n\_sces}$^2$} & $^3\mathrm{R}_\mathrm{vir}$&$^4\mathrm{M}_\mathrm{vir}$              &  \multicolumn{2}{c}{LIRGs$^5$} \\
       &      &$(\prime^2)$ &  (5$\sigma$)&        min.$^1$       &      C    &         T             &                            &                                         &       Obs       &      Exp     \\
       &      &             &   \multicolumn{2}{c}{($\mu$Jy)}     &           &                       &                            &                                         &                 &              \\
\hline \noalign{\smallskip}
CL0024 & 0.39 & 37.8        &  \  140     &         141   \       & 13        &         35            & 0.94                       & 6.42                                    & 10              & -            \\
\noalign{\smallskip}
A370   & 0.37 & 40.5        &  \  350     &         208   \       &  1        &         20            & 0.91                       & 5.53                                    & 1               & 8            \\
\noalign{\smallskip}
A1689  & 0.18 & 36.0        &  \  450     &         320   \       & 11        &         18            & 1.1                        & 5.7                                     & 0               & 1            \\
\noalign{\smallskip}
A2390  & 0.23 & 7.0         &  \  100     &         \ 54  \       &  4        &         28            & 1.62                       & 20.35                                   & 0               & 1            \\
\noalign{\smallskip}
A2218  & 0.18 & 20.5        &  \  125     &         \ 90  \       &  6        &         46            & 1.63                       & 18.27                                   & 0               & 1            \\
\noalign{\smallskip}
\multicolumn{11}{l}{$^1$Faintest source considered in publication.}                                                \\
\multicolumn{11}{l}{$^2$Number of cluster sources, and total number of IR sources.}                                \\
\multicolumn{11}{l}{$^3$Cluster virial radius\,($\mathrm{h}^{-1}$\,Mpc)}                                             \\
\multicolumn{11}{l}{$^4$Cluster virial mass\,($\mathrm{h}^{-1}$\,$10^{14}$\,$M_{\odot}$)}                            \\
\multicolumn{11}{l}{$^5$Number of LIRGs (or near LIRGs $(L_{IR}\,>\,9\,\times\,10^{10}\,L_{\odot}$) detected vs.}  \\
\multicolumn{11}{l}{\ \ number expected if cluster were to be similar to CL0024+1654.}                             \\
     \hline
      \end{tabular}
\end{center}
\end{table*}

The two most distant clusters observed with ISO are CL0024+1654
($z=0.39$, Coia et al. 2004) and J1888.16CL ($z=0.56$, Duc et al. 2004).
These are among the deepest ISOCAM observations and could 
detect several cluster members. (For CL0024+1654, 13 out of 35 sources found at 15\,$\mu$m are
spectroscopically confirmed to be cluster sources. For J1888.16CL, 6 out of 44 sources found 
at 15\,$\mu$m are so confirmed.)  A common feature of these two clusters is the high star
formation rate inferred from their MIR luminosities.  These two
observations were also the most extended cluster maps performed in terms of absolute cluster
area covered at the cluster, $2.3\,\times\,2.3$
and $1.3\,\times\,6$ square Mpc for CL0024+1654 and J1888.16CL, respectively.
This fact, and evolutionary effects detectable around $z\,\sim\,0.5$ (see
Dressler et al.\,1999), may explain the high IR luminosity of the 15\,$\mu$m
sources found in these clusters.

In the case of J1888.16CL, Duc et al.\,(2004) estimate star-formation rates
ranging between 20 and 120\,$M_{\odot}$ per year. At least six galaxies
belong to the cluster and have IR luminosities above
$1.3\times\,10^{11}L_{\odot}$. In CL0024+1654, Coia et al.\,(2004b) report
ten sources brighter than $9\,\times\,10^{10}L_{\odot}$. The star formation rates inferred from
the MIR flux are one to two orders of magnitudes greater than those
based on the [OII] flux (though in this case the comparison was only possible for the three sources for
which [OII] data was available.) 
This is compatible with the result in A1689 (Fig.4) and 
implies similar dust extinction characteristics.

Interestingly, the galaxies emitting at 15$\mu$m appear to have a
spatial distribution and a velocity dispersion slightly different from the
other cluster galaxies.  Galaxies in CL0024+1654 are detected
preferentially at larger radii, with the velocity dispersion of
15$\mu$m sources being greater than that of the galaxies in the cluster.
In J1888.16CL,
Duc et al. (2004) estimate that to explain the number of sources
detected on the basis of infall of galaxies from the field an infall rate
of about 100 massive galaxies per 100\,Myr is required, which seems unrealistic.
Numerical simulations
and X--ray observations show however that accretion onto clusters from
the field is not a spherically symmetric process, but occurs along
filaments or via mergers with other groups and clusters.  One
therefore cannot exclude the possibility that the LIRGs observed in
these distant clusters belonged to such a recently accreted structure. An
alternative possibility is that the collision with an accreted group
of galaxies stimulated star formation in the galaxies of the group as
a consequence of a rapidly varying tidal field (Bekki 1999).
This could be the case for CL0024+1654 and A1689, clusters which
show evidence of accreting groups of galaxies in their multi-modal
velocity distributions. CL0024+1654 is in the process of interacting with
a smaller cluster.



\subsection{A diffuse intra-cluster dust component ?}
\label{sec:environment}

The hot intra-cluster material contains metals, and so is not entirely 
primordial. Might not the stars which produced the metals also have deposited dust
in the ICM? The first to note that emission from intra-cluster material might 
be observable were Yahil \& Ostriker (1973), based on a galactic dust-to-gas
ratio and the observed intra-cluster gas.  Ostriker \&
Silk (1973) and Silk \& Burke (1974) developed expressions for the
lifetime of dust in a hot intra-cluster medium.  Pustilnik (1975),
drawing upon contemporaneous reports of optical absorption in
clusters, attributed it to dust and estimated that cluster
emission at 100\,$\mu$m would be in the $10^3$ to $10^4$\,Jy range for
6 nearby clusters.  Voshchinnikov \& Khersonskii (1984) also 
attributed claimed reddenning of galaxies in distant clusters to dust
absorption, and estimated that the total FIR emission from the Coma or
Perseus clusters should be $10^5$ to $10^6$\,Jy (tens of Jy/arcminute$^2$) 
in the 50 to 100\,$\mu$m range. They
estimated the sputtering lifetime of intra-cluster dust grains to be
up to $10^8$\,years. Hu et al. (1985), noting that intra-cluster dust must be short-lived, 
predicted FIR dust emission of a few Jy per square
degree, close to the IRAS limit, for a sample of X-ray luminous clusters.

IRAS measurements failed to bear out even the most modest of the 
above predictions. Kelly \& Rieke (1990) co-added IRAS scans across 71 clusters with $0.3
\leq z \leq 0.92$ to arrive at an average 60\,$\mu$m value for cluster emission
of 26$\pm$5 mJy per cluster, and 46$\pm$22\,mJy at 100\,$\mu$m.
Dwek et al. (1990) refined models of intra-cluster dust and its
interactions and calculated an upper limit of 0.2~MJy/sr for dust-emission
from the Coma cluster, consistent with IRAS observations.  Then total cluster
emission would not be more than a few Jy at the peak wavelength
(around 100 $\mu$m). They concluded that dust in the cluster centre
could not explain the visual extinction, nor could cluster galaxies or
their halos.  Dust in the outskirts could, if it were un-depleted. But
they saw no mechanism for the production of such dust.
Wise et al. (1993) analysed 56 clusters at 60 and 100 $\mu$m 
from clusters with a range of X-ray emission, and some without
cDs. For the only two clusters (A262 and A2670) showing a far-infrared 
excess lacking an immediate explanation (in terms of point sources or cirrus) they concluded
that the result was likely to be due to discrete sources in the clusters.  Averaged over the
sample as a whole there was evidence of excess FIR at the 2-$\sigma$
level. No large FIR excesses associated with cooling flows were found.
Bregman et al. (1990) looked for evidence of star formation in
27\,cD galaxies. In half of their sample of X-ray-bright clusters
they found IR, X-ray and blue luminosities to be comparable,
consistent  with dust grains heated by the X-ray emitting gas,
thereby suggesting that dust cooling can compare with thermal
bremsstrahlung as a cooling mechanism for the intra-cluster gas. 

Cox et al. (1995) studied a much larger sample of 158 Abell clusters,
again at 60 and 100 $\mu$m, and after making a more rigorous
correction for spurious sources due to galactic cirrus, they concluded
that only about 10\% of cD galaxies in rich clusters have significant
FIR emission, but with luminosities ten times greater than the X-ray
luminosities produced in the cores of clusters, a condition which they
therefore regarded as transient for any individual cluster. If the FIR
emission comes from dust heated by the intra-cluster thermal electrons,
significant dust sputtering is expected on timescales of several
10$^8$ years. Dust must then be replenished to account for continuous
IR emission, presumably through the mechanisms discussed for stripping
material from cluster galaxies (see Section~\ref{sec:intro}).

By the launch of ISO one could imagine a ``life-cycle" of dust in a
cluster, tracing the flow of material - gas and dust - out of
infalling galaxies, the destruction of dust in the high-temperature
intra-cluster medium, and its possible eventual re-deposition,
through the mechanism of cooling flows, into the cluster-dominant
galaxies. But only upper-limits or occasionally, and with marginal
significance, global or average cluster FIR emission, could constrain
scenarios for the role of dust in the physics of the clusters as a
whole.

Stickel et al. (1998, 2002) used ISOPHOT to observe extended FIR
emission of six Abell clusters. Strip scanning measurements were 
performed at 120\,$\mu$m and 180\,$\mu$m. The raw profiles of the
I$_{120\,{\mu}m}$/I$_{180\,{\mu}m}$ surface brightness ratio including
zodiacal light show a bump towards Abell 1656 (Coma), dips towards
Abell 262 and Abell 2670, and are without clear structure towards
Abell 400, Abell 496, and Abell 4038. After subtraction of the
zodiacal light and allowance for cirrus emission, only the bump 
towards Abell 1656 (Coma) is still present (Fig.\,6). This excess of
$\approx\,0.2$\,MJy/sr seen at 120\,$\mu$m towards Abell 1656 (Coma)
is interpreted as thermal emission from intra-cluster dust
distributed in the hot X-ray emitting Coma intra-cluster medium. The integrated excess
flux within the central region of 10$^{\prime}$ to 15$^{\prime}$
diameter is $\sim$\,2.8\,Jy. Since the dust temperature is poorly
constrained only a rough estimate of the associated dust mass of
$M_D\,\sim\,10^7\,M_{\odot}$ can be derived. The associated visual
extinction is negligible (A$_V$\,$\ll$\,0.1\,mag) and much smaller
than claimed from optical observations. No evidence is found for
intra-cluster dust in the other five clusters observed. 

\begin{figure}
\resizebox{0.5\hsize}{!}{\includegraphics{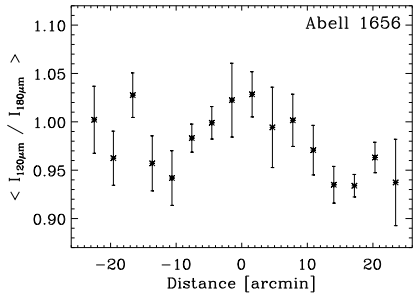}}
\caption{From Stickel et al. (2002): The overall zodiacal-light-subtracted surface
brightness ratio I$_{120\,{\mu}m}$/I$_{180\,{\mu}m}$ for A1656 (Coma) averaged
over both scan position angles and all detector pixels.}
\end{figure}

Quillen et al. (1999) suggested integrated
emission from the cluster galaxies as the most likely source for the
detected signal at the centre of Coma. Stickel et al. (2002) replied
that if this was indeed the case, the same signal should have been
detected in {\em all} clusters observed.

The absence of any signature for intra-cluster dust in five clusters and the
rather low inferred dust mass in Abell 1656 indicates that intra-cluster dust is
probably not responsible for the excess X-ray absorption reported in
cooling flow clusters (White et al. 1991).  These observations thus 
represent a further unsuccessful attempt to detect the presumed
final stage of the cooling flow material. This agrees with numerous
previous studies at other wavelengths, while spectroscopic observations 
with ESA's XMM-Newton X-ray observatory have shown that intra-cluster gas does not 
cool beyond $\sim 1$~keV (see, e.g., Molendi \& Pizzolato 2001), so dust 
deposition in cooling flows is not actually expected. Not surprisingly, 
then, further attempts by Hansen et al. (1999, 2000a, b) also failed 
to detect dust associated with the cooling flows presumed to exist at the
centres of most galaxy clusters.

\section{Conclusions}
\label{sec:conclusions}

Building upon the legacy of IRAS, ISO could consolidate a number of
important fundamenatal conclusions about the properties of galaxies in
the Virgo, Coma and other nearby galaxy clusters.  The correlation between
galaxy Hubble type and mid- and far-infrared properties was firmly 
established. A major new cold dust component was identified by ISOPHOT 
observations, which could not have been found by IRAS.  Moderate resolution
FIR spectroscopy was possible for Virgo cluster galaxies using LWS, establishing
a correlation between the strength of the [CII] line and the FIR flux and a
two order of magnitude difference in [CII] to near-IR ratio 
between early type galaxies and late (spiral) types.
Mid-infrared emission (5 to 18\,$\mu$m) was found to correlate with star
formation, but to trace it less faithfully when star formation rates become
high enough for UV photons to disrupt the infrared-emitting materials.
However, the MIR emission traces well the FIR and bolometric emission.
In general the properties of galaxies in nearby clusters ($z\,<\,0.1$) were
found to exhibit little dependence on the cluster environment.

ISO, for the first time, could extend mid-infrared observations to
clusters beyond z\,=\,0.1, and in so doing has detected over 100
galaxies in clusters in the redshift range 0.17 to almost 0.6. 
Although the collected observations on seven  such clusters were
rather heterogeneous, a number of important trends  are found in
the ISO results. There is a clear tendency for clusters at higher
redshift to exhibit higher average rates of star formation in their
galaxies, and numerous LIRGs have been observed in such clusters.
There is tentative evidence for an association between the infrared
luminosities found and galaxy infall from the field, but substantial
evidence to link high levels of star-formation in cluster galaxies
to the dynamical status of a cluster, and to interactions with other
(sub-)clusters. It seems clear that star formation rates
deduced for cluster galaxies from optical tracers often fall one to
two orders of magnitude below rates derived from MIR emission levels, 
so that star formation in cluster galaxies may have been seriously
underestimated in some cases, in the past.

ISO found little evidence for widespread infrared emission from dust in the 
intra-cluster medium.

Clearly, what ISO has not been able to supply has been systematic
large area mapping of a substantial sample of galaxy clusters out
to high redshift. Observing times to achieve such coverage with ISO
would have been prohibitive, given the multi-position and 
heavily-overlapped rasters that would have been required. If galaxy
evolution within clusters is to be further explored and related to
galaxy evolution in the field, clusters of different masses and
dynamical status must be studied systematically in the MIR and FIR
out to well over 1 virial radius in order to understand how and
why the IR properties of  galaxies change from cluster to cluster,
and from cluster to field. Large  area coverage is important since
it is known that significant modifications  of the galaxy
properties already occur in the outskirts of galaxy  clusters
(G\'omez et al. 2003; Kodama et al. 2001; Lewis et al. 2002).

Several programmes underway or planned with the Spitzer Space Observatory
should thoroughly address these challenges.


\end{document}